# Experimental demonstration of OSC at IOTA: IOTA Run #3 Report

J. Jarvis*[1], V. Lebedev*[1], A. Romanov[1], D. Broemmelsiek[1], K. Carlson[1], S. Chattopadhyay[1,2,3], A. Dick[2], D. Edstrom[1], I. Lobach[4], S. Nagaitsev[1,4], H. Piekarz[1], P. Piot[2,5,] J. Ruan[1], J. Santucci[1], G. Stancari[1], A. Valishev[1]

[1] Fermi National Accelerator Laboratory, Batavia, IL, USA.
[2] Department of Physics, Northern Illinois University, DeKalb, IL, USA.
[3] SLAC National Accelerator Laboratory, Stanford University, CA, USA.
[4] Department of Physics, The University of Chicago, Chicago, IL, USA.
[5] Argonne National Laboratory, Argonne, IL, USA.

## Introduction

Optical Stochastic Cooling (OSC) is an optical-bandwidth extension of Stochastic Cooling that could advance the state-of-the-art cooling rate in beam cooling by three to four orders of magnitude [1-3]. The concept of OSC was first suggested in the early 1990s by Zolotorev, Zholents and Mikhailichenko, and replaced the microwave hardware of SC with optical analogs, such as wigglers and optical amplifiers. A number of variations on the original OSC concept have been proposed, and while a variety of proof-of-principle demonstrations and operational uses have been considered, the concept was not experimentally demonstrated up to now [4-9]. An OSC R&D program has been underway at IOTA for the past several years [4]. Run #3 of the IOTA ring, which began in Nov. 2020 and concluded in Aug. 2021, was focused on the world's first experimental demonstration of OSC. The experimental program was successful in demonstrating and characterizing the OSC physics with the major outcomes including strong cooling in one, two and three dimensions, validation of the theoretical models of OSC and the demonstration of OSC with a single electron. This report briefly describes the activities and major milestones of the OSC program during Run #3. Detailed descriptions of the OSC theory, conceptual design and hardware elements can be found in reference [4].

## OSC Program

The Run #3 OSC program was divided into three phases:

- Apparatus Commissioning (APP_COMM)
- Demonstration Experiment (DEMO_1)
- Systematic Studies of OSC Concepts (SSOCs).

### Ph1: Apparatus Commissioning (APP_COMM)

*Objectives: Installation of the OSC apparatus, injection and lattice correction, validation of all critical diagnostic and control systems*

Installation of the OSC hardware began on 10/09/20. The original E-straight hardware and BR nonlinear insert were removed from the ring and secured elsewhere in the enclosure, and a temporary replacement spool for BR was installed. The modules of the OSC insertion were installed, starting from the M4 main dipoles and working inwards towards the OSC bypass. Installation of the OSC vacuum envelope, which included the in-vacuum optics and the precision-motion systems, was completed on 12/05/21, and full functionality of the motion systems was verified shortly thereafter. The completed OSC system is pictured in Figure 1. Some effort was required to detect and eliminate vacuum leaks on button flanges of the BPM in the ER OSC spool. A minor leak was also detected in the OSC optics cross but intervention was not required as the ultimate pressure there was comparable to the rest of the ring.

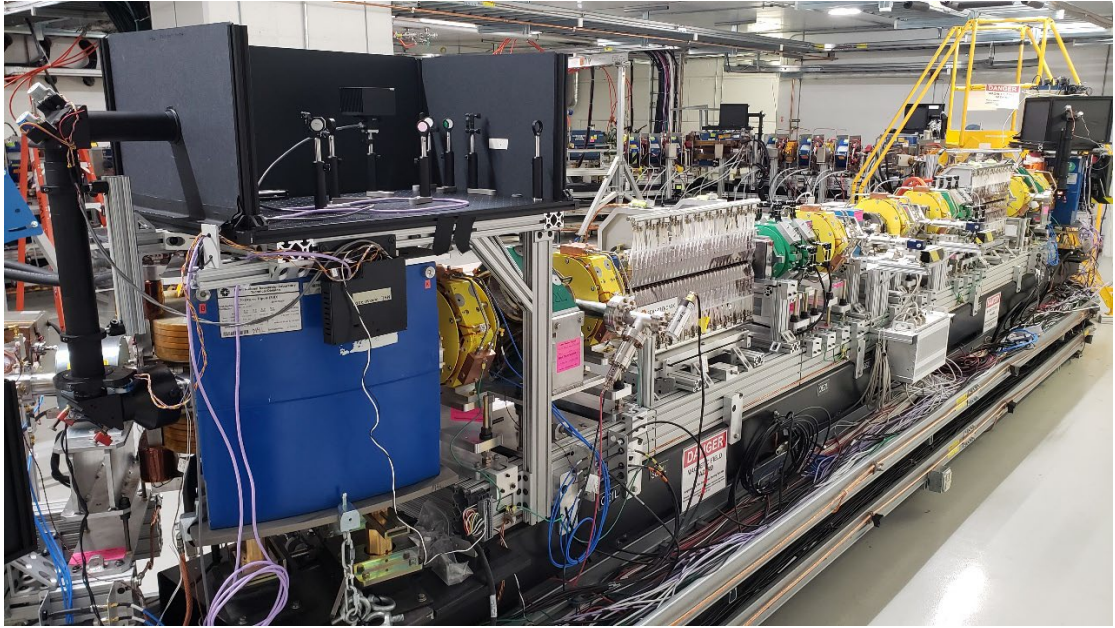

**Figure 1:** OSC apparatus installed in IOTA's E straight.

During this same period, all IOTA ring sextupoles were installed in the appropriate locations and rough aligned. During installation, it was noted that the vacuum spools in the CL/CR are not mirror symmetric across the short axis (east-west) of the ring. The known interference issue with the sextupoles in the CL/CR straights means that the short/long sextupoles in these sections are also not symmetric. The short-yoke version is located downstream in both sections.

Initially, temporary power supplies with standard regulation (ZUPs) were used for the OSC dipoles and quadrupoles, and a single turn was made in IOTA on 12/22/21. After the holiday break, various downtime incidents, and the installation of the final magnetic elements (vertical correctors and OSC coupling quad), three turns were made on 02/11/21, and stored beam was achieved on 02/15/21. The performance of the lattice was continually improved throughout the run. Due to initial difficulties with the low-compaction (~1.7e-3) OSC lattice, a higher-compaction (~4.9e-3) lattice was developed and used for the duration of the run. A variety of adjustments were made to the chamber positions to improve beam aperture and lifetime. Systematic (beam) aperture studies showed ~80% of theoretical maximum. Ultimately, maximum injections of ~0.3 mA were achieved and low-current lifetimes >20 minutes. Using measurements of the beam size in the corrected lattice, the ultimate pressure was estimated to be ~$3.7 \cdot 10^{-8}$ Torr of atomic hydrogen equivalent, which coincides with the vacuum estimate of the previous IOTA Run to within ~15% accuracy. Additional improvements in pressure would require the IOTA bakeout system, which is being installed in advance of Run #4.

During commissioning, the M2R station was upgraded for higher resolution using two hardware improvements: the rejection of vertically polarized light (Thorlabs PBSW-405) and the use of a narrow band filter (Thorlabs FBH405-10). Although the emitted vertically polarized light is weak, it increases the diffraction-limited spot size by ~20%. The wavelength of the narrowband filter was made as short as possible while still maintaining a high quantum efficiency in the camera's sensor. The filter reduced the diffraction contribution of long-wavelength radiation and the contribution of lens chromaticity from short-wavelength radiation. Calculations indicate that these measures reduced the diffraction contribution by almost a factor of two. To focus the beam image, the longitudinal position of each camera was adjusted to minimize the measured beam size; preference was given to the vertical size as depth-of-field effects are more significant in the horizontal size due to the horizontal curvature of the beam trajectory. The theoretical diffraction-limited spot size for M2R is ~15 μm.

The M4L diagnostics station and OSC laser alignment system (LAS) were installed in early March. Figure 2 shows the basic layout of the M4L lightbox, including the "fundamental line" and SPAD detector that were later used for single-electron OSC experiments. The LAS comprises a HeNe laser (outside M4R), which is aligned through two surveyed pinholes (outside M4R and M4L), and matching optics that focus the beam near the center of the PU. The laser then naturally relays through all OSC optics and diagnostic lines to produce an absolute reference for detector and beam alignment; an example of the focused LAS spots on the KU01/02 cameras is shown in Figure 2. The transverse alignment of the LAS relative to IOTA's coordinate system should be on the order of tens of microns in position and tens of microradians in angle. The supplies that normally power QE2R/L were temporarily connected to the OSC undulators to enable initial testing of undulator diagnostics and the alignment system. Voltage limits of these supplies restricted the undulators to a maximum current of ~7 A, which corresponded to on-axis fundamental wavelength of ~632 nm. Narrowband filters were used to improve visibility of the undulator radiation (UR) against the radiation from the main dipoles. The focused "first light" from the undulators on KU02 is shown in Figure 3.

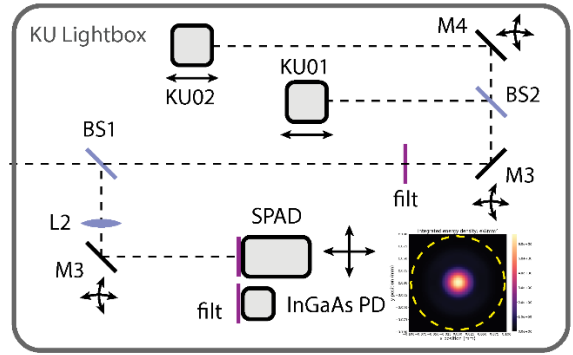
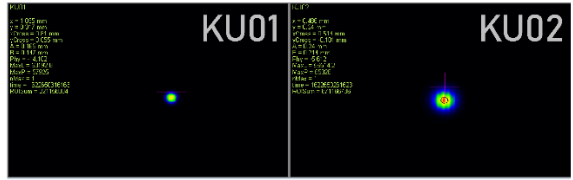

**Figure 2:** (top) Schematic of the M4L diagnostics station. The inset heatmap shows the simulated focused fundamental UR on the SPAD detector, and the yellow dashed line represents the detector's active area. (bottom) Focused LAS spot on the KU01 and KU02 cameras.

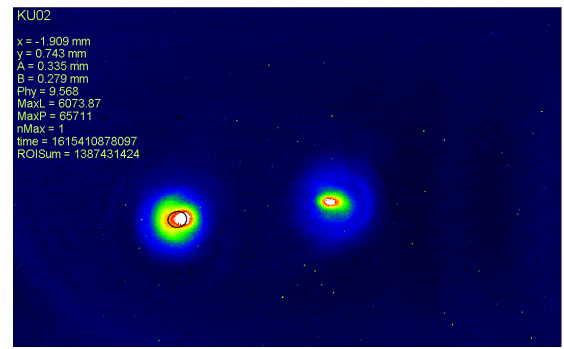

**Figure 3:** First undulator light in the OSC apparatus. The PU and KU light are displaced from one another by translation of the in-vacuum lens.

Lattice correction, aperture studies and diagnostics upgrades continued throughout March while preparations were being made for the main cable pull (~03/31/21). The cable pull was a prerequisite for delivering full current to the undulators, high-quality power regulation (1e-5) to the OSC bypass and control signals to the M3R streak camera (to be installed in May). Following the cable pull, the BiRa PCRC current regulators were connected to the OSC chicane dipoles and attempts were made to observe interference between the PU and KU radiation. The narrowband filter mentioned above relaxed the accuracy requirements of the absolute delay setting, and on 04/07/21, interference of the UR was observed for the first time. The relative delay of the beam and light was scanned using two methods: rotation of one of the optical delay plates and sweeping the excitation of the chicane dipoles. The modulated KU+PU intensity on KU02 recorded during the scan, and the periodicity of the modulations was in excellent agreement with models of the delay stage and the OSC lattice (specifically the path lengthening in the OSC bypass vs chicane strength as computed in 6dsim); the results are shown in Figure 4.

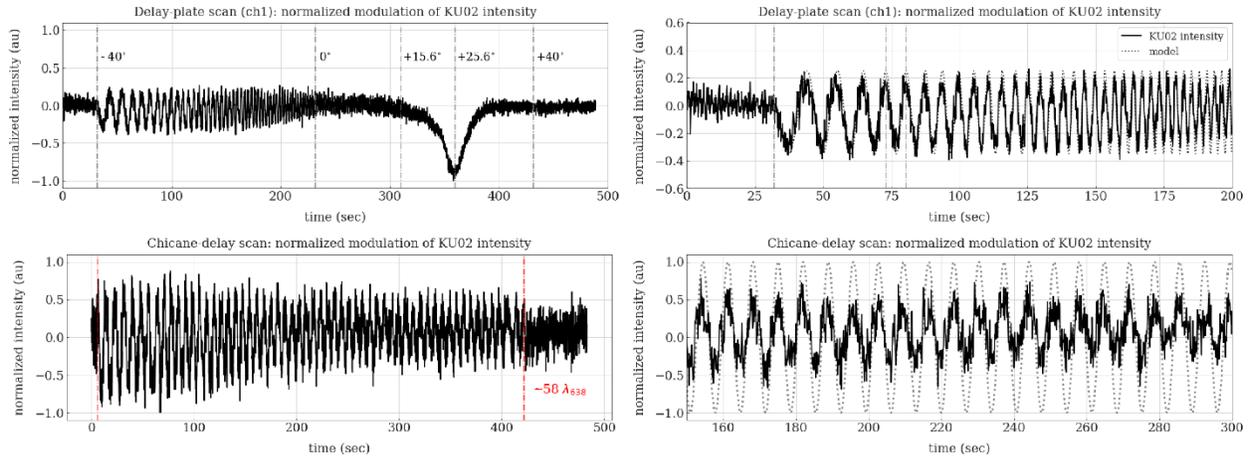

**Figure 4:** Intensity modulation of the KU+PU fundamental UR (632nm) during delay-plate (top) and chicane strength (bottom) scans. Left: full scans; Right: zoomed with model fits. For the excitation range of the chicane scan, the exact number of expected modulations were observed, and the modulation frequency agreed with the model to a few percent.

Following additional lattice-correction studies, an inspection of the IOTA main quadrupoles was performed, and significant gaps (~0.5-1 mm) were discovered between the halves of two of the AmPS-type quadrupoles (QA4R and QC2R). The gaps were eliminated, and additional lattice correction was performed. For more precise longitudinal alignment, the 632-nm filter was removed so that the coherence length of the UR was set by the bandwidth of the UR fundamental. The observation of interference in this configuration on 04/16/21 signaled successful longitudinal alignment of the OSC system. The undulators were then connected to their 10-kW power supplies, enabling nominal excitation at 48A (950-nm fundamental). The undulators were briefly brought to full power on 04/16/21. At this point, all the essential systems required for a first attempt at OSC were operational. The last major diagnostic systems, the streak camera and SPAD, were brought online during phases 2 and 3 of the program, respectively, and are discussed in the next sections.

### Ph2: Demonstration Experiment (DEMO_1)

*Objectives: Alignment of OSC systems and observation of effect of OSC; optimization of strength and characterization of essential parameters*

The Blackfly-PGE-23S6M-C CMOS cameras used to image the UR have a quantum efficiency (QE) of ~5% percent at 950 nm, which enables direct imaging of the UR fundamental. A 900-nm longpass filter was inserted in the KU02/KU01 to reduce contamination from UR harmonics and the M4L and chicane dipoles. A representative image of the focused UR on the KU01 (upstream) and KU02 (downstream) cameras is shown in Figure 5. Two in-vacuum lenses were used during the OSC program. An error by the manufacturer of the first lens resulted in a focal length that was ~5% too long. The radiation from the PU was then underfocused, particularly at the upstream end of the KU. This is clearly seen in Figure 5 (top row), especially when compared to the result for a properly manufactured lens (bottom row), which was installed in the final weeks of the program. SRW simulations of the radiation field experienced by the particles for these two cases are also presented in Figure 5.

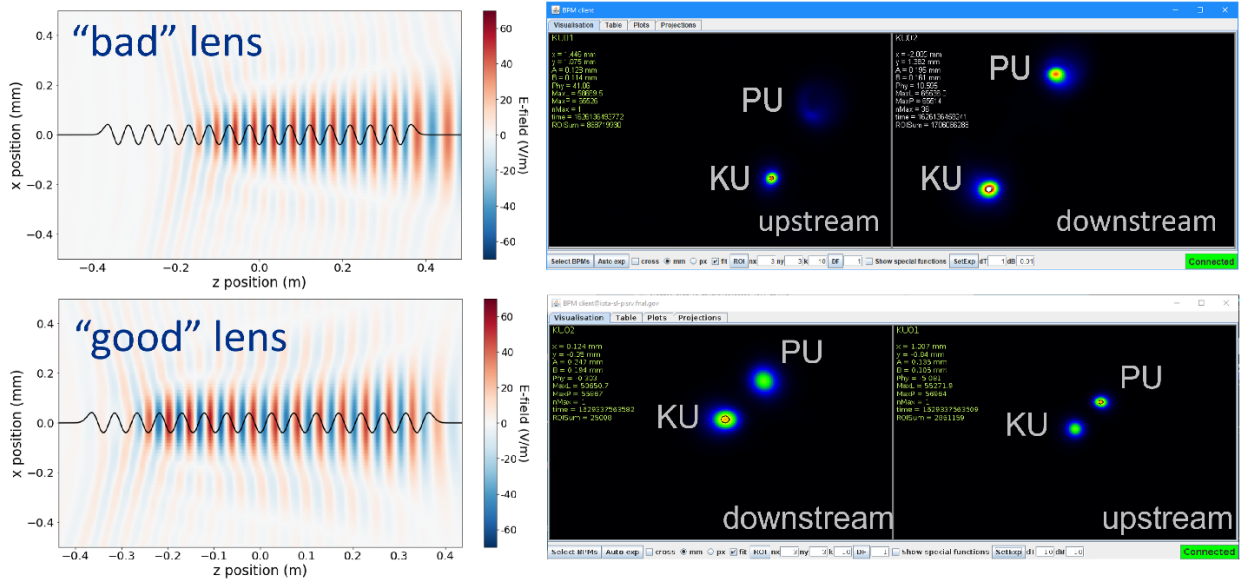

**Figure 6:** Focused UR spots from the PU and KU on the KU01 (upstream) and KU02 (downstream) cameras for two different lenses: "bad" (top; +5% focal-length error) and "good" (bottom; nominal focal length). The "good" lens was installed with ~2 weeks remaining in the run.

On 04/20/21 the undulators were brought to full power, the PU and KU spots were overlapped using lens translations and closed-orbit (CO) bumps, and the interference zone was quickly located near the nominal delay settings that were used for 632 nm. The interference effect was then maximized using additional CO bumps and lens translations. Strong, global modifications of the beam distribution were then observed, indicating successful observation of the underlying OSC physics. Figure 6 shows the integrated KU+PU intensity on KU02 during a full scan of the delay system through the interference zone. When first observed, the interfering UR fluctuated wildly with a ~1-second timescale, and the beam distribution also transitioned between heated and cooled states with a similar frequency.

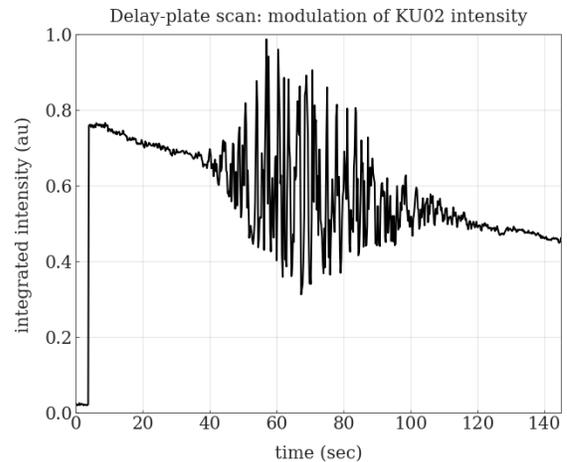

**Figure 5:** Integrated intensity of the interfering UR on KU02. The large fluctuations described in the text are clearly visible during the scan.

A malfunction in the cryogenics system on 04/21/21 resulted in the suspension of OSC operations until 05/05/21. During this period, a Hamamatsu model C5680 dual-sweep streak camera with a synchroscan (M5675) vertical deflection unit was installed for measurement of the beam's longitudinal distribution in the OSC experiments. The system was installed on the M3R diagnostics station (Figure 7) after appropriate modifications to the support structure. A 50/50 non-polarizing beam splitter was used to direct half of the SR from the existing M3R SR BPM to the entrance slit of the streak camera, and a Blackfly-PGE-23S6M-C CMOS camera was used as the detector element, and. An external clock generator was phase locked to IOTA's 4th-harmonic RF (30 MHz) and used to drive the streak-camera's sweep at the 11th harmonic (82.5 MHz) of the beam's circulation frequency (7.5 MHz). To calibrate the streak camera image (ps/pixel), IOTA's RF voltage was first calibrated using a wall current monitor to measure shifts in the synchronous phase of the beam for different voltage settings. Measurements of the synchrotron frequency (via resonant excitation of the beam) were then made as a function of voltage setting, which yielded a small correction factor for the momentum compaction (+15%). This value is highly sensitive to focusing errors in the low emittance lattice designed for the OSC studies. Finally, the streak camera's calibration factor was determined by fitting the measured longitudinal beam position as a function of voltage. The first images of the beam's longitudinal distribution were acquired on 05/07/21.

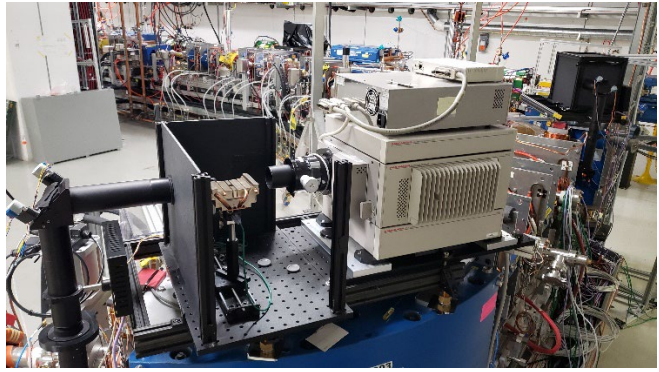

**Figure 7:** Streak camera installed on the modified M3R diagnostics station.

On 05/13/21, the Main Injector (MI) ramp was identified as the principal cause of the instability observed in the OSC system. The periodicity of the modulations in the interfering UR at KU02 matched that of the MI ramp (MIBEND). The regulation of IOTA's main bend supply (IBEND) was upgraded, and the OSC effect was stabilized. An example of stabilized UR interference (from 07/15/21) is given in Figure 8. The first well-focused images from the streak camera were acquired on 05/14/21 and with additional tuning of the OSC system, strong longitudinal cooling and heating were observed on 05/17/21. Examples of the heating, cooling and off states are given in Figure 9.

In the following week, the fundamental-diagnostics line, shown the bottom half of the M4L station schematic (Figure 2) was installed in preparation for single-electron experiments and fast measurements of OSC damping. On 05/27/21, the IOTA SR BPM software was upgraded for live plotting of the rms fits and python scripts for automated delay scanning were written. These two upgrades enabled meaningful systematic tuning of OSC, an example of which is given in Figure 10.

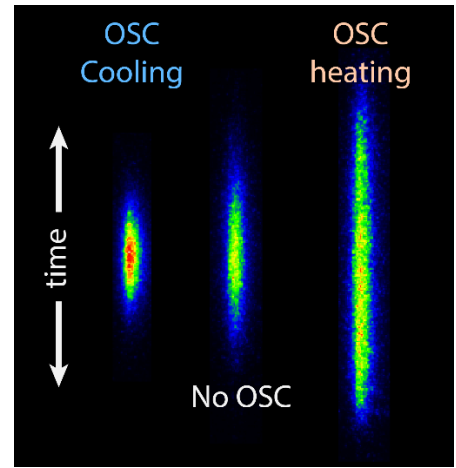

**Figure 8:** Early examples of the beam's longitudinal distribution in the OSC-cooling, OSC-off and OSC-heating modes.

Extensive tuning and characterization of OSC was carried out during the months of June and July. Initial attempts were made to observe the damping with "fast" measurements using a SPAD at M4R, but the results of these initial tests were difficult to interpret. One of the most important observations from this operations period was the complexity of OSC behavior when coupling was included in the system. The fully uncoupled case (longitudinal only) was relatively straightforward to tune and seemed qualitatively consistent with expectations from theory; however, when the cooling was coupled to the horizontal plane, and especially when the lattice was operated on a transverse coupling resonance, the cooling and heating between the transverse and longitudinal planes would often appear out of phase. An example of this effect during OSC toggles is given in Figure 11. This was thought to be connected to subtle misalignments of the beam and light within the KU and perhaps related to the poor structure of the underfocused PU radiation. The observation of significant interaction with the 2nd-harmonic UR on the low-delay side of the scans was also suggestive of misalignment, as the even harmonics should have zero strength "on axis" in a properly aligned system.

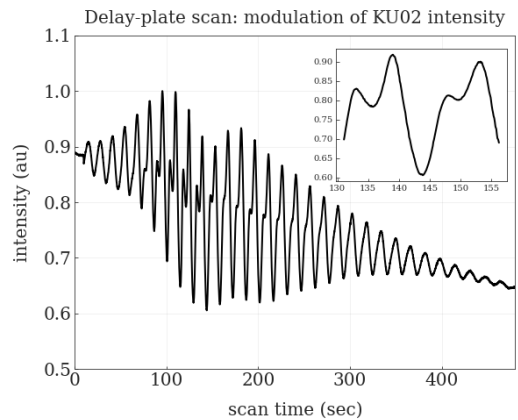

**Figure 9:** Integrated intensity of the interfering UR on KU02 after regulation upgrades to IBEND. The inset shows a detailed view of the strongest OSC zone. The distortions that occur on the heating side of each zone likely result from imperfect overlap of the PU and KU coherent modes for high synchrotron amplitudes.

Within the limitations of the available correctors, exhaustive alignment studies were performed in which the CO was bumped in the two undulators and the lens position was finely tuned. An unexpected horizontal-to-longitudinal (delay) coupling was also observed during manipulations of both the CO and the lens. As a result, effective tuning required long, slow delay scans for every set point OSC system, which significantly increased the difficulty of tuning. The alignment (CO and lens position) of the full OSC system was rebuilt multiple times and studies were performed to better understand the impact of the undulator trims on the CO.

In this phase, the OSC physics was conclusively demonstrated for the first time; however, despite extensive tuning, system configurations supporting simultaneous cooling of the longitudinal and transverse planes had not yet been achieved. OSC operations were suspended from 07/24/21 to 08/05/21 for vacations and the commissioning of a 150 MeV test lattice for the nonlinear integrable optics program.

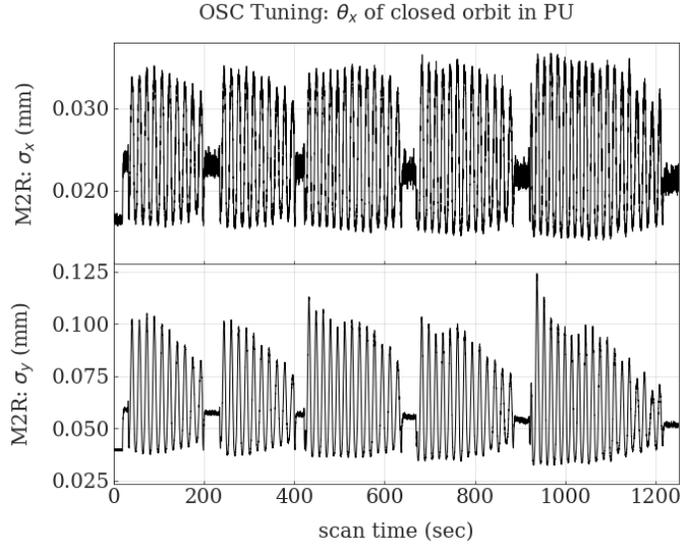

**Figure 10:** Example of OSC tuning scans. In this example, each scan was performed for a different amount of $\theta_x$ bump in the PU.

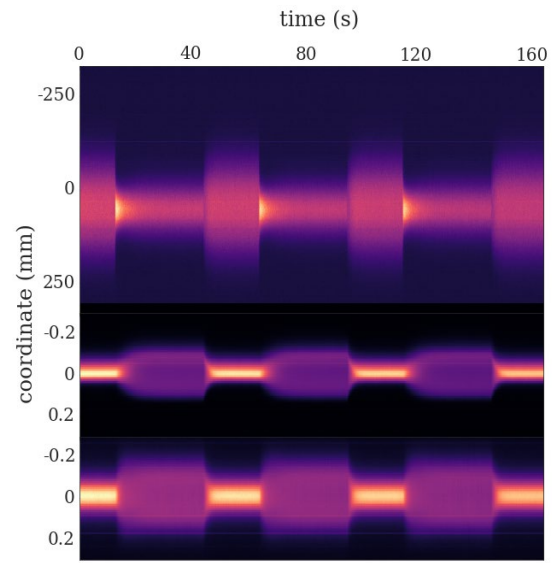

**Figure 11:** OSC cooling toggles showing antiphased behavior of longitudinal (top) and transverse (middle: x, bottom: y) planes during longitudinal cooling. STREAK camera intensity falls at the start of cooling due to excitation of particles outside of the acceptance of the streak camera's input aperture.

**Ph3: Systematic Studies of OSC Concepts (SSOCs)**

*Objectives: Optimized configurations for 1D, 2D and 3D; full characterization of OSC performance (e.g. rates & ranges) in different configs and regimes; single-electron OSC experiments*

Shortly after the resumption of operations, stable 2D (s,x) and 1D (s) cooling were achieved (08/10/21 and 08/12/21). While 1D cooling should correspond to zero excitation of the coupling quad, excitation at approximately half the design value (~0.5 A) was required to null the coupling to the transverse planes. This was a first indication that there may be additional terms in the transverse-to-longitudinal mapping of the OSC bypass. Having collected representative data for all system configurations using the first in-vacuum lens, a vacuum intervention was made on 08/13/21, and the in-vacuum lens was replaced with one having the proper focal length. A dry-nitrogen purge and continuous flow throughout the procedure enabled a rapid pump down of the system. Stored beam with meaningful lifetime (several minutes) was achieved within three days. The bottom panel in Figure 5 illustrates the improved focusing of the PU radiation using the new lens. On 08/16/21, first OSC with the new in-vacuum lens was achieved. OSC tuning and performance improved significantly with the new lens.

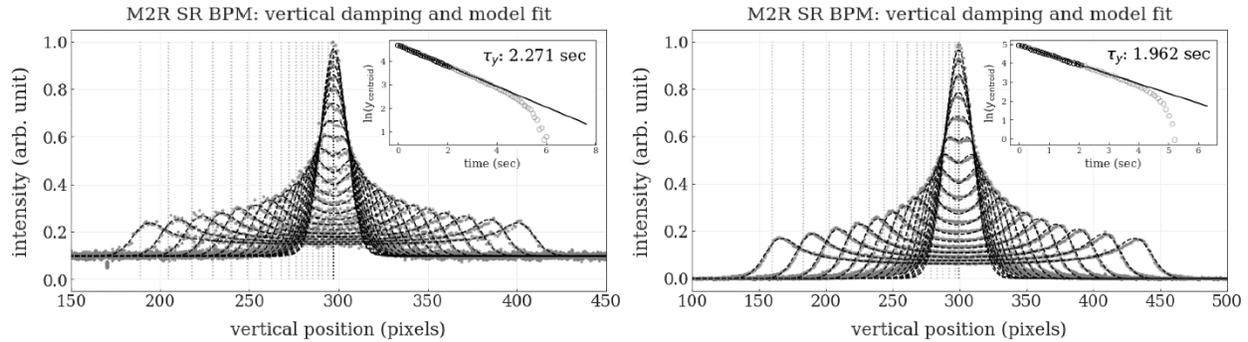

**Figure 12:** Projected vertical distribution on M2R after a ping with IOTA's vertical kicker in the original, ~97-MeV lattice (left) and the scaled, ~101-MeV lattice (right). Prior to the measurements, the tunes were split to xy uncouple the lattice. Insets show the measured damping time due to SR damping.

With the new lens installed, and having recovered OSC in initial testing, studies began for finalizing the lattice and OSC-system configurations for final OSC measurements. At the design energy of 100 MeV, the fundamental UR and the SPAD QE should have adequate overlap for OSC experiments; however, in Ph2, the maximum achieved SPAD signal (~200 Hz using a non-polarizing beam splitter at M4L and ~1000 Hz using a 900 +nm dichroic) in the fundamental line was more than an order of magnitude below the expected value (~10 kHz with the dichroic). Measurements of vertical SR damping with the lattice xy uncoupled indicated a beam energy of ~97 MeV, several percent below the design value and consistent with the significantly reduced SPAD signal. On 08/17/21, the excitation of IBEND, ILAM, the chicane and all quads were raised by ~4% with a target energy of ~101 MeV, and the lattice and CO were corrected as before. Data and fits for the energy measurements are presented in Figure 12. On 08/18/21, using the new 101-MeV lattice, 3D OSC was demonstrated for the first time.

LOCO data with the new lattice indicated the presence of significant vertical dispersion (~6 cm) in the OSC bypass. Due to the lack of skew quad correctors in the region, this could only be addressed by physical rotation of the OSC quads. Model fits indicated that rolls about the beam axis on the on the order of 0.4° were required for several of the OSC quads. Dial gauges and a precision digital level were used to ensure the accuracy of the applied rotations. The procedure was successful in reducing the vertical dispersion by a factor of ~20 (3 mm). This was the first LOCO-based physical update (rotation or translation) made in IOTA.

After final lattice and orbit corrections, full system configurations (CO and lens tuning) were built for each OSC mode (1D, 2D, 3D). OSC was carefully optimized in each case and ancillary data such as tunes measurements and lifetime data were collected. OSC scans were then systematically performed for the optimized configurations. Table 1 shows a summary of all OSC measurements that were performed in the final lattice configuration with large numbers of particles ($>10^4$). Data sets include delay scans and cooling/heating toggles, and each scan includes an OSC cooling toggle at the beginning and end. CAMDAQ hardware servers were not well synchronized for some sets. In such cases, the discrete nature of the toggles can be used to align the data. For delay scans, the delay plates were swept over the given angular interval (see Elog for precise readback values) at a constant rate of 0.01°/s, which corresponds to approximately one wavelength every 30 seconds. In all cases, raw images were taken at intervals of 0.5 seconds. The data itself comprises timestamped raw images ('rawz' format) for the listed cameras. Numerous other data sets were taken in the form of projections throughout Ph2 and Ph3 of the program. Full details can be found in the FAST Elog. A representative delay scan and toggle for the 3D OSC case are shown in Figures 13 and 14, respectively.

**Table 1:** Summary table of the final data sets for optimized OSC in each configuration. Data sets include delay scans and cooling/heating toggles. Each scan includes an OSC-cooling toggle at the end. For delay scans, the delay plates were swept over the given angular interval (see Elog for precise readback values) at a constant rate of 0.01°/s. In all cases, raw images were taken at 0.5-s intervals; SCAN06-2Qx is has the coupling quad at twice the nominal excitation.

| Scan ID | Shift date | OSC type | IBEAMS (μA) | Cams. | ~[$\theta_i,\theta_f$] (deg) | $\theta_{cool}$ |
|---|---|---|---|---|---|---|
| SCAN03 | 8/24/2021 | 1D | 0.65 | M2R, M1L, STREAK | [50.6602,47.4463] | 49.0532 |
| SCAN04 | 8/24/2021 | 1D | 0.2 | M2R, M1L, STREAK | [50.6602,47.4463] | 49.0532 |
| SCAN05 | 8/24/2021 | 1D | 0.1 | M2R, M1L, STREAK | [50.6602,47.4463] | 49.0532 |
| SCAN06 | 8/24/2021 | 1D | 0.05 | M2R, M1L, STREAK | [50.6602,47.4463] | 49.0532 |
| SCAN04 | 8/25/2021 | 3D | 1.8 | M2R, M1L, STREAK | [51.4959,48.2819] | 49.8888 |
| SCAN05 | 8/25/2021 | 3D | 0.85 | M2R, M1L, STREAK, KU02 | [51.4959,48.2819] | 49.8888 |
| SCAN06 | 8/25/2021 | 3D | 0.2 | M2R, M1L, STREAK, KU02 | [51.4959,48.2819] | 49.8888 |
| SCAN07 | 8/25/2021 | 3D | 0.1 | M2R, M1L, STREAK, KU02 | [53.1029,43.4609] | 49.8888 |
| TOGGLES08 | 8/25/2021 | 3D | 0.05 | M2R, M1L, STREAK | - | - |
| SCAN04 | 8/26/2021 | 2D | 0.85 | M2R, M1L, STREAK, KU02 | [51.0909,47.8770] | 49.4839 |
| SCAN05 | 8/26/2021 | 2D | 0.4 | M2R, M1L, STREAK, KU02 | [51.0909,47.8770] | 49.4839 |
| SCAN06 | 8/26/2021 | 2D | 0.2 | M2R, M1L, STREAK, KU02 | [51.0909,47.8770] | 49.4839 |
| SCAN07 | 8/26/2021 | 2D | 0.1 | M2R, M1L, STREAK, KU02 | [52.6980,43.0560] | 49.484 |
| TOGGLES08 | 8/26/2021 | 2D | 0.05 | M1L,STREAK | - | - |
| SCAN06-2Qx | 8/27/2021 | 2D | 0.15 | M1L, STREAK | - | - |

The principal features of the projections in Figure 13 can be well understood within the existing theoretical framework. At the beginning and end of the scan (high and low delay, respectively), the particles and light are longitudinally separated in the KU, which effectively turns off OSC and results in the equilibrium state being set by SR damping alone. As the scan proceeds, OSC alternates between cooling and heating modes with the total number of modulation periods being approximately twice the number of undulator periods. The OSC strength peaks after about ten periods due to intentional overfocusing by the in-vacuum lens, and the strength falls off to either side according to the bandwidth of the integrated system. In the heating mode, large particle amplitudes in one plane can lead to an inversion of OSC in the other planes. This is clearly seen in Figure 13b (white dashed line) where strong longitudinal antidamping results in cooling for the transverse planes despite the OSC system being tuned for the heating mode. The effect is less apparent for the horizontal plane than for the vertical due to the large (dispersive) contribution of the momentum spread to the horizontal beam size at the SR monitor's location. In contrast, when OSC is weak and comparable to SR damping, as it is toward the edges of the scan (see Fig. 13a), the antidamping is weak, and the cooling and heating are fully synchronized across the different planes.

Analysis of the rms fits indicates a total cooling force of ~18 s$^{-1}$, which is approximately half of the value anticipated from SRW simulations. A 3D scan of the integrated apparatus was made using a Creaform Metrascan3D metrology system and was compared with the mechanical model of the apparatus. The results indicate a few-mm misalignment of one of the bypass chambers, which may restrict the physical aperture for the PU light. SRW simulations that include this aperture restriction yield an expected reduction of the total cooling force on the order of 15%. Other potential sources of loss may include nonlinearities in the bypass mapping, the possibility of distorted CO trajectories in the undulators due to saturation in the steel poles, and reduced energy exchange due to the finite radiation spot size in the KU. Analysis of the beam sizes indicates an equilibrium emittance (SR damping only) that is four-times larger than the nominal value and consistent with scattering from residual-gas molecules. The larger beam size would exacerbate any OSC losses due to the finite radiation spot size. The inferred gas pressure agrees with similar analyses from IOTA Run#2 to within ~15%.

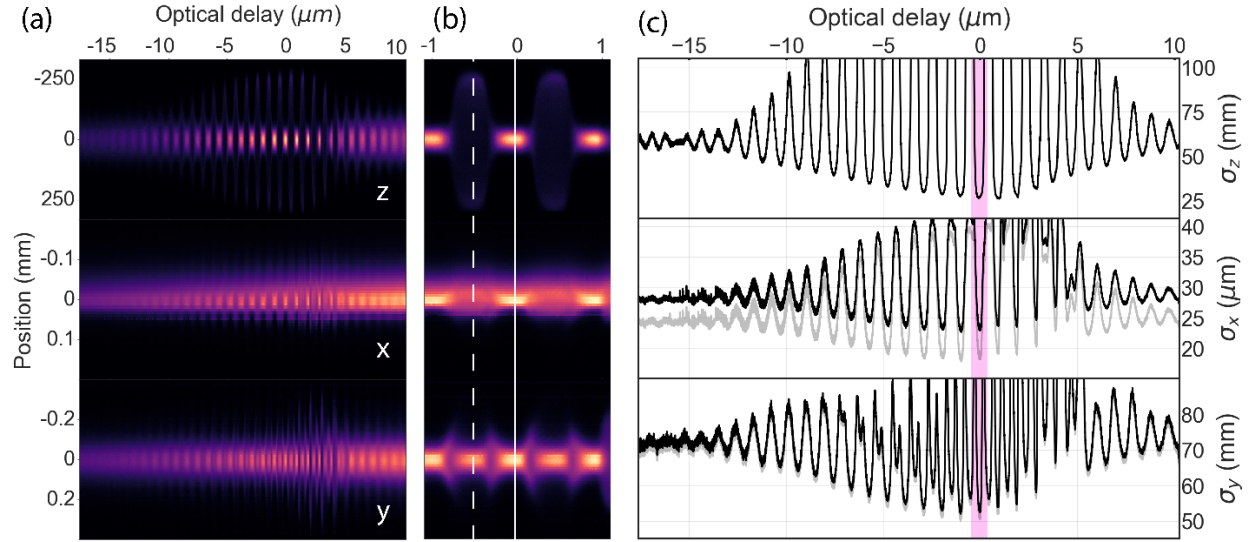

**Figure 14:** Projected beam distributions for a delay scan in the 3D OSC configuration As measured (*z,x,y*) projections of the beam distribution during an OSC delay scan. The transverse (*x,y*) distributions were recorded on the M2R SR monitor. The decay of the average intensity over the ~15-min scan is consistent with the beam's natural 1/e lifetime (~17 min) due to scattering with residual gas. **b,** Zoomed view near the strongest OSC cooling (vertical solid line) and heating (dashed line) zones. The intensity for each projection is renormalized for clarity. **c,** Rms beam sizes for the projected distributions. The fit is performed on the Gaussian core of the beam to reduce the impact of depth of field effects in the horizontal plane and to avoid contamination by the non-Gaussian tails that result from gas scattering. For all planes, the vertical axis is clipped at the amplitude where the core of the distribution becomes non-Gaussian due to OSC heating. Diffraction-corrected curves are shown in gray, and the pink band marks the location of the strongest OSC cooling zone. The number of modulations in the beam size is approximately twice the number of undulator periods, as expected from theory.

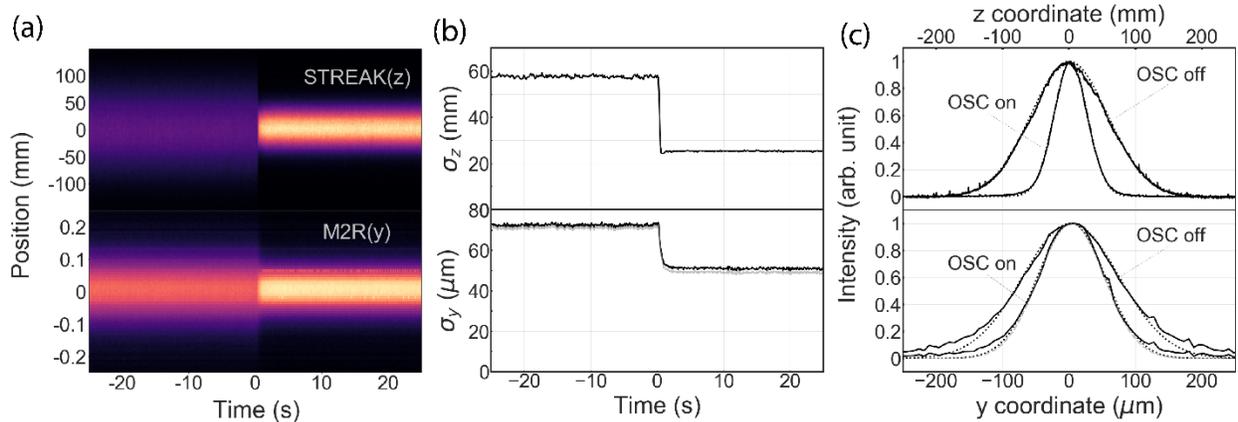

**Figure 13:** Fast toggle of the OSC system. **a,** Dependence on time of single dimensional beam distributions in *z* (streak camera) and *y* (M2R SR monitor) during an OSC toggle. The system is initially detuned by $30\lambda$ and is snapped to the maximum cooling setting at *t*=0. **b,** Rms beam sizes from Gaussian fits of the raw projections presented in **a**. **c,** Distributions averaged over time (solid lines) and their Gaussian fits (dotted lines) for the OSC-off and OSC-on states for the intervals of [-20,-10]s and [10,20]s. In **b** and **c**, the M2R fits use only the central +/-125 μm to reduce contamination by the non-gaussian tails resulting from gas scattering. Diffraction-corrected curves are shown in gray, and the distributions in each case have been normalized to a peak value of one for comparison.

Analysis of 3D-OSC projections for the heating mode (e.g. Figure 15) yields estimates of the longitudinal cooling range that are in good agreement with theory. The equilibrium amplitudes were computed using two separate approaches: from the ratio of OSC to SR damping rates and directly from the streak-camera calibration. For the maximum achieved OSC rates, the computed amplitudes coincide with one another to ~5% accuracy. In Figure 13c, the longitudinal distribution is well fit by a pure Gaussian, indicating that the beam is well within the longitudinal cooling range and there is no reduction of the damping rate for the distribution tails. Data taken for the 1D-OSC configuration at very high currents (~10 μA) in the 1D OSC configuration, such as those shown in Figure 16, show indications of particle accumulation at the 2$^{nd}$-order cooling zone.

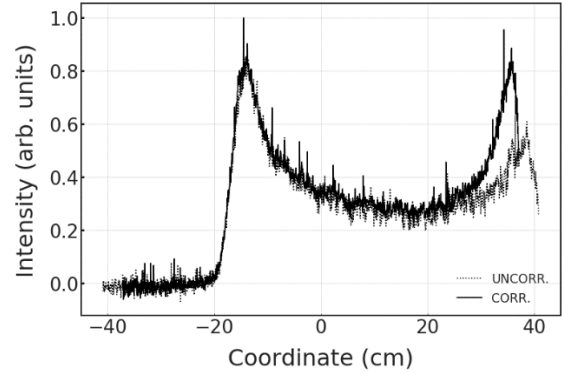

**Figure 15:** Correction of nonlinearity in streak camera image. Longitudinal distribution for OSC in the antidamping mode before (dotted) and after (solid) correction of non-linearity in the streak camera image.

The transverse cooling range could not be studied in the 3D data due to the relatively weak transverse OSC, which precluded particle trapping at large transverse amplitudes; however, we note that the measured rms transverse mode emittances of the cooled beam at the OSC maximum are ~0.9 nm, which is almost 100-times smaller than the expected cooling acceptance and at least ~30-times smaller than the worst-case estimate of the cooling acceptance when reduced by the nonlinearity of longitudinal displacements at large amplitudes. Therefore, cooling-range limitations are not expected to play any role in the OSC measurements. Analysis of data for the 2D-OSC configuration, especially the case with double excitation of the coupling quad, should yield estimates of the transverse cooling range for comparison with theory. Figure 17 presents an example of 2D-OSC heating with simultaneous accumulation of particles at separate synchrotron and betatron attractors. The 2D data at higher coupling strength also show an inversion of the longitudinal OSC force while in the heating mode due to large betatron amplitudes. This is the complement to the inversion of transverse OSC seen in the 3D data of Figure 14.

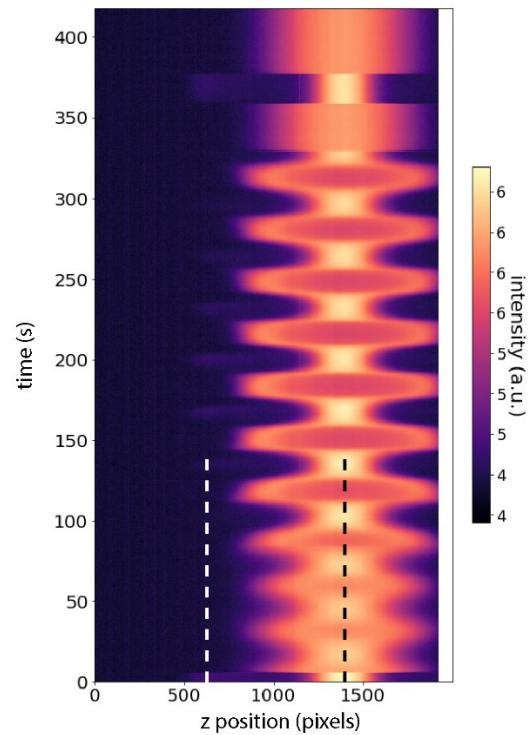

**Figure 16:** STREAK projections for a delay scan in the 1D OSC configuration with high beam current (~10 μA). The 2$^{nd}$-order cooling zone can be seen as particle accumulation at high amplitude (white dashed line) while tuned for the cooling mode. The intensity map is shown on a log scale to increase visibility of the 2$^{nd}$-order zone.

A full analysis of representative data from the various OSC configurations is ongoing.

All measurements described above are based on observation of the equilibrium distributions under different conditions, rather than direct "fast" measurement of the damping. On the final day of operations (08/28/21), the IOTA SR BPM software was upgraded to enable high-speed projections (~200 Hz) from a single camera with a reduced region of interest. A longitudinal kicker system was quickly assembled and tested. The kicker system, activated by a clock event, phase modulated IOTA's main RF generator near the synchrotron frequency for four synchrotron periods. Projections were recorded with and without 1D OSC for a variety of excitation amplitudes. Figure 18 shows an example comparison of SR damping only and 1D-OSC (tuned for max cooling) for a moderate amplitude kick. Full analysis of the fast-damping data set is pending.

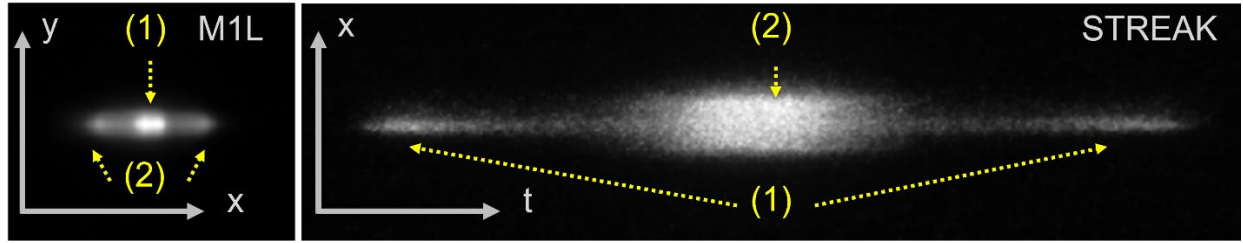

**Figure 18:** Example beam distributions for 2D-OSC heating measured on the M1L SR monitor (left) and STREAK (right). These simultaneous images clearly show the accumulation of particles at two attractors, one with large synchrotron amplitude (1) and one with large betatron amplitude (2).

The final program element for Ph3 was the demonstration and characterization of single-electron OSC. On 08/26/21, the fundamental line in the M4L lightbox was realigned and systematic scans were used to locate and maximize the SPAD signal. The maximum signal (count rate) achieved was ~8 kHz per electron, which is consistent with expectations. Modulation of the SPAD signal due to UR interference was immediately observed during initial delay scans with a depth of modulation at the ~75% level.

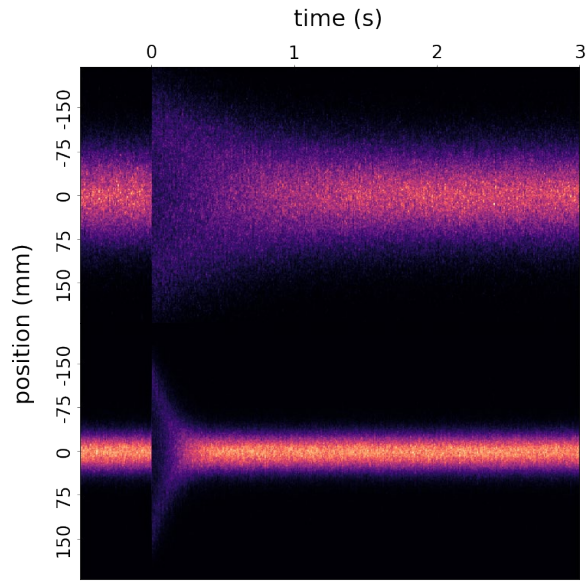

**Figure 17:** Example fast measurement of damping with (bottom) and without (top) 1D OSC after excitation by the longitudinal kicker system.

Systematic studies were performed over a two-day period collecting delay scans and toggles for the 1D and 2D-OSC configurations. The collected data consists of raw images of the transverse distribution on the M1L SR monitor and photo-multiplier tube (M3L) and SPAD timing data collected with the Hydraharp event timer. Figure 19 shows an example of binned timing data for the PMT and SPAD during a 1D-OSC toggle. Although the temporal resolution of the integrated system is insufficient to resolve the difference between the equilibrium sizes for the OSC-on and OSC-off cases, the effect of the strong OSC damping is clearly visible in the suppression of the large synchrotron excitations resulting from gas scattering. Turning points are visible in the heating mode as well as brief intervals where OSC is effectively turned off/reduced by the excitation of large betatron amplitudes (via gas scattering). Examples of slow (~0.03λ/s) and fast (~15λ/s) delay scans are given in Figure 20. Slow scans display the same structure as those with large particle number, including distortion of the interference in the OSC heating zones, whereas fast scans show the interference effects without substantial modification of the particle motion.

Figure 21 presents an example of a 2D-OSC toggle for a single electron. The longitudinal data are substantially similar to those in Figure 19; however, the particle's behavior is dramatically different when the system is tuned for the heating zone. When combined with the M1L SR-monitor data, also shown in Figure 21, it is apparent that the particle is transitioning spontaneously between the betatron and synchrotron attractors discussed previously (Figure 18). Figure 22 presents averaged raw images of the particle SR at M1L for the two bistable states. For the large-synchrotron-amplitude case, the dispersive contribution to the distribution is apparent. The transverse cooling range can be extracted from the large-betatron-amplitude case and compared to the "beam" measurements described previously. Finally, as expected for the 2D-OSC case, excitations in the $y$ plane appear unaffected by the state of the OSC system, while those in the $x$ plane are strongly suppressed. Finally,

**Additional Information:**

Three primary publications are in progress: 1) announcement of the successful demonstration of OSC, 2) comprehensive description and analysis of all OSC experiments with "beam", 3) OSC experiments with a single electron.

The Run#3 OSC data and various supporting files are stored on the IOTA-FAST network drive:

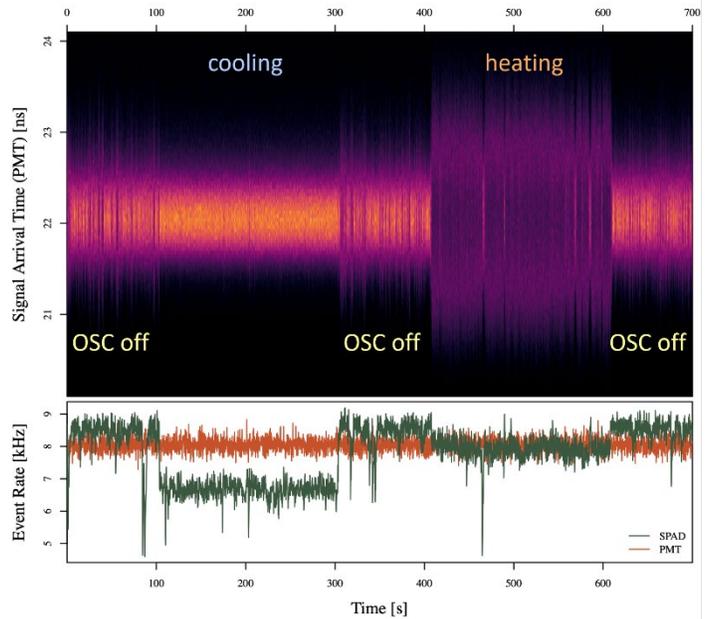

**Figure 19:** Binned PMT/SPAD timing data for a 1D-OSC cooling/heating toggle. (top) PMT data showing the suppression of synchrotron excitations in the cooling mode and turning points in the heating mode. (bottom) Total PMT and SPAD counts during the toggle series. The RF voltage was lowered from the nominal setting of ~110V to ~55V to increase the natural "bunch" length and thus improve the visibility of the heating-mode turning points.

\\beamssrv1.fnal.gov\iota-fast.bd\Data\Run_3\OSC

All OSC data have also been backed up on a physical drive.

The IOTA OSC redmine site provides additional links and program details and is located at https://cdcvs.fnal.gov/redmine/projects/iota-osc/wiki. The day-to-day activities and operations of the program were documented in detail on the FAST Elog.

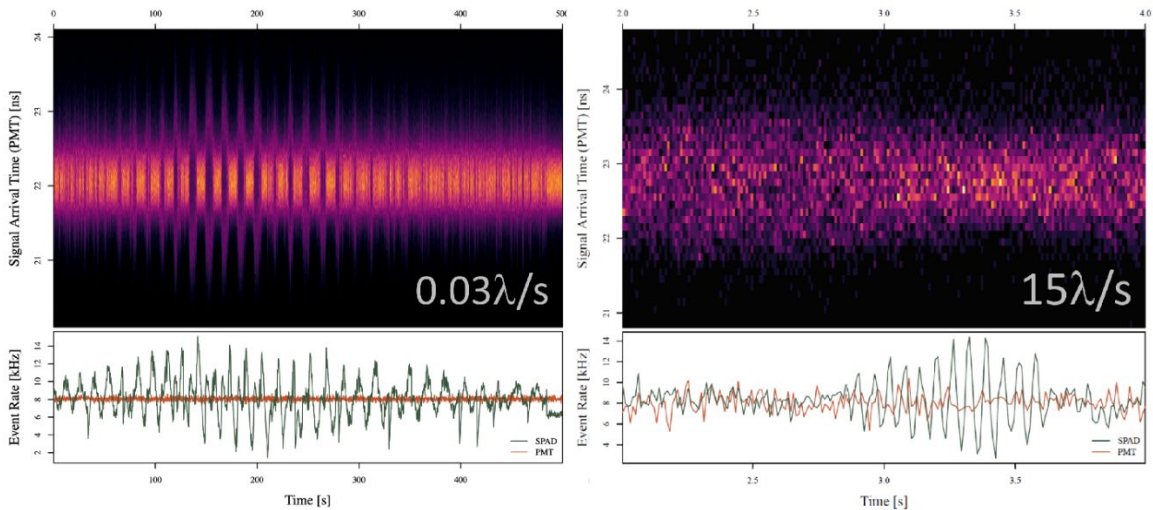

**Figure 20:** Slow (left) and fast (right) 1D-OSC delay scans for a single electron.

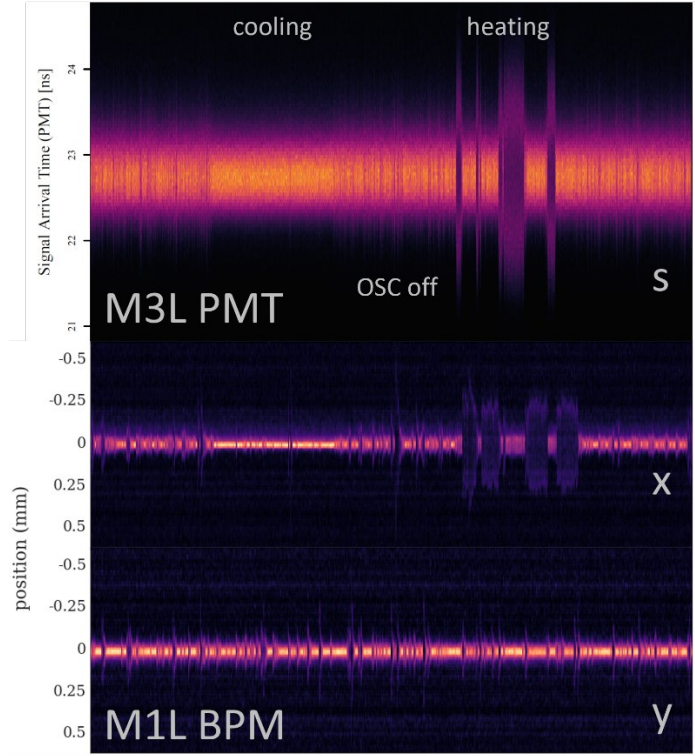
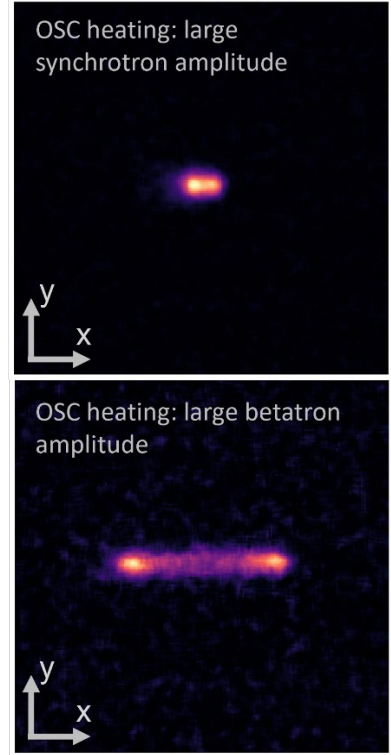

**Figure 22:** Binned PMT/SPAD timing data for a 2D-OSC cooling/heating toggle. (top) PMT data showing the suppression of synchrotron excitations in the cooling mode and turning points and bistable behavior in the heating mode. (middle and bottom) M1L SR monitor projections during the scan. Transitions between the two OSC attractors is apparent.

**Figure 21:** Transverse distributions on the M1L SR monitor for the two bistable states: (top) large synchrotron amplitude and (bottom) large betatron amplitude.

**Acknowledgements:**

We are grateful to D. Frank, M. Obrycki, R. Espinoza, N. Eddy and J. You for extensive technical and hardware support. We also thank B. Cathey for operations support and acknowledge useful discussions with M. Zolotorev, M. Andorf, A. Lumpkin, J. Wurtele, A. Charman and G. Penn. This manuscript has been authored by Fermi Research Alliance, LLC under Contract No. DE-AC02-07CH11359 with the U.S. Department of Energy Office of Science, Office of High Energy Physics. This work was also supported by the U.S. Department of Energy Contract No. DE-SC0018656 with Northern Illinois University and by the U.S. National Science Foundation under award PHY-1549132, the Center for Bright Beams.